Microscopy AND
Microanalysis

# Automated Grain Boundary Detection for Bright-Field Transmission Electron Microscopy Images via U-Net

Matthew J. Patrick[1] , James K. Eckstein[2] , Javier R. Lopez[3] , Silvia Toderas[1] ,
Sarah A. Asher[1] , Sylvia I. Whang[4] , Stacey Levine[5] , Jeffrey M. Rickman[6] ,
and Katayun Barmak[1],*

[1]Department of Applied Physics and Applied Mathematics, Columbia University, 200 S.W. Mudd Building, 500 W. 120 Street, New York, NY 10027, USA
[2]Department of Physics, University of Illinois, 1110 W. Green Street, Urbana, IL 61801, USA
[3]Department of Mechanical Engineering, Columbia University, 210 S.W. Mudd Building, 500 W. 120 Street, New York, NY 10027, USA
[4]Department of Physics, Barnard College, 504A Altschul Hall, 3009 Broadway, New York, NY 10027, USA
[5]Department of Mathematics and Computer Science, Duquesne University, 440 College Hall, 1100 Locust Street, Pittsburgh, PA 15282, USA
[6]Department of Materials Science and Engineering, Lehigh University, Whitaker Lab 244, 5 E. Packer Avenue, Bethlehem, PA 18015, USA
*Corresponding author: Katayun Barmak, Email: kb2612@columbia.edu

## Abstract

Quantification of microstructures is crucial for understanding processing–structure and structure–property relationships in polycrystalline materials. Delineating grain boundaries in bright-field transmission electron micrographs, however, is challenging due to complex diffraction contrast in images. Conventional edge detection algorithms are inadequate; instead, manual tracing is usually required. This study demonstrates the first successful machine learning approach for grain boundary detection in bright-field transmission electron micrographs. The proposed methodology uses a U-Net convolutional neural network trained on carefully constructed data from bright-field images and hand tracings available from prior studies, combined with targeted postprocessing algorithms to preserve features of interest. The image processing pipeline accurately estimates grain boundary positions, avoiding segmentation in regions with intragrain contrast and identifying low-contrast boundaries. Our approach is validated by directly comparing microstructural markers (i.e., grain centroids) identified in U-Net predictions with those identified in hand tracings; furthermore, the grain size distributions obtained from the two techniques show notable overlap when compared using *t*-test, Kolmogorov–Smirnov test, and Cramér–von Mises test. The technique is then successfully applied to interpret new microstructures having different image characteristics from the training data, with preliminary results from platinum and palladium microstructures presented, highlighting the versatility of our approach for grain boundary identification in bright-field micrographs.

## Introduction

The microstructural length scale of nanocrystalline thin films necessitates the use of the transmission electron microscope (TEM) in order to resolve the details of the grain structure. Imaging of such samples is frequently done in the bright-field (BF) mode, with an example shown in Figure 1 for two aluminum (Al) films with columnar grains that span the thickness of the sample. Though the grain structure of the film comprises relatively simple polygonal shapes, the complex contrast inherent in these BF images renders quantitative microstructural analysis particularly challenging.

These interpretive challenges have their origins in the physics of the imaging technique. BF TEM leverages diffraction contrast, meaning that different crystal orientations will show different intensities in the final image depending on the fraction of the beam that is diffracted and the fraction that is directly transmitted through an aperture. The technique is sensitive to small, local changes in orientation leading to intragrain contrast, such as bend contours that complicate image interpretation. A further complication is that two adjacent grains in a polycrystal may have an identical diffracted intensity, showing no contrast between them, and so, in some contexts, no visible grain boundary.

For these reasons, automated approaches to grain boundary detection in BF TEM images have made little if any additional progress (Chambers, 2006) since the work reported by Carpenter et al. (1998), in spite of successful implementations of automated grain boundary identification in other imaging modes (Nakane et al., 2016; Campbell et al., 2017; Gupta et al., 2020). Consequently, observation of grains in nanocrystalline thin films and the extraction of microstructural data, like the extensive geometric and topological data reported in Barmak et al. (2013), have relied on labor-intensive, manual analysis to trace the grain boundaries. This requires many person-hours of tedious work to obtain data for a statistically significant population of grains (typically ~1,000 grains), severely limiting throughput.

The grain structure of polycrystalline thin films can also be investigated using crystal orientation mapping via precession electron diffraction (PED). Using this method, with appropriate cleanup and postprocessing parameters, mean grain size and grain size distributions in nanocrystalline thin films have been obtained (Liu et al., 2014; Rohrer et al., 2017) and acquisition and the significant amount of time and expertise needed for reliable manual labeling of training data (Xu et al., 2022). Additionally, previous applications of neural networks to microstructural image analysis have not encountered the challenging contrast issues found in BF TEM image data (Li et al., 2020; Perera et al., 2021; Jung et al., 2022), therefore preventing the transferability of previous approaches. Nevertheless, the nature of BF TEM image data, accompanied by expert hand-traced grain boundaries, renders the automated grain boundary identification problem ripe for data-driven machine learning approaches. As no standard data sets are available for training in this problem, this work instead leverages a large hand-labeled data set available from prior work in our laboratory to overcome these hurdles; this data set is now publicly available via the link listed in the Availability of Data and Materials section.

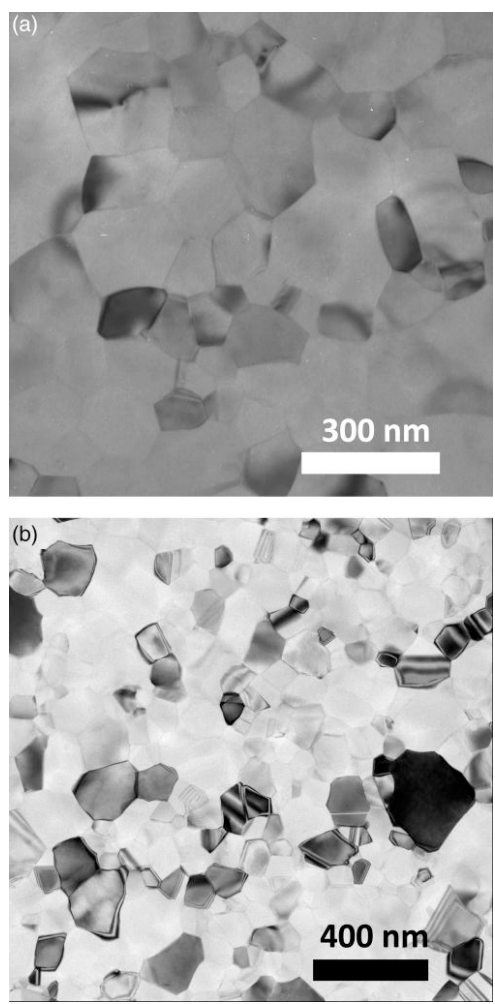

**Fig. 1.** (**a**) Scanned and inverted photographic negative of a BF TEM image of a 100-nm-thick Al film, representative of training and validation data used in this work. (**b**) BF TEM image, collected on a CCD camera from a recently deposited 100-nm-thick Al film, representative of test for this work.

The grain boundary detection problem is similar to, but distinct from, the image segmentation problem, in which foreground objects are separated from the underlying background regions (Li et al., 2020; Perera et al., 2021). CNNs are proven, powerful solutions for the segmentation problem, with one such CNN's architecture, known as U-Net, was originally developed for the automated identification of cellular structures in BF TEM images of stained biological samples (Ronnenberger et al., 2015). Image segmentation algorithms, including U-Net, are usually concerned with identifying distinct, separate objects dispersed or clustered throughout a FoV, with thick boundaries, empty space between them, or regions with distinct characteristics (Ronnenberger et al., 2015; Xu et al., 2019; Tran et al., 2018; Horwath et al., 2020; Kirillov et al., 2023). Indeed, previous attempts at using U-Net for segmenting grains in diffraction contrast images of polycrystalline materials have shown preliminary success in identifying bright grains in dark-field images but fail to resolve boundaries between them in clusters and fail to reproduce microstructures from BF images (Xu et al., 2022). Furthermore, the typical use of these algorithms in biomedical contexts is not the resolution of the fine features of a network of cell borders. In contrast, the grain boundary identification problem can be seen from two complementary perspectives; in both cases, the details of the boundaries must be resolved. From the first perspective, the problem focuses on the identification of grains, space filling objects that cover the entire FoV and are separated by exceptionally narrow boundaries, while, from the second perspective, the focus is on the identification of grain boundaries that constitute a dense network after binarization. In both cases, segmentation requires special attention to training data set construction and postprocessing as simple thresholds do not successfully reproduce the microstructure from model predictions.

However, BF TEM images of nanocrystalline thin films captured via diffraction contrast and BF images of cells captured using mass–thickness contrast have some shared challenges in their processing. While the physics of imaging is fundamentally different in the two modes, both have intrafeature contrast and biological samples still often contain touching or adjacent objects of the same class. Furthermore, due to similarly laborious labeling, both problems suffer from limited training data sets. U-Net and its offshoots have been demonstrated to overcome these issues in biological and materials samples (Ronnenberger et al., 2015; Punn and Arawal, 2020; Xu et al., 2019; Zhou et al. 2018; Sadre et al., 2021), making the architecture attractive for grain boundary detection in BF TEM micrographs. This is especially true, in light of the observation that neural networks trained on microstructural have shown good agreement with the distributions obtained by hand tracings. Unlike BF imaging, PED-based crystal orientation mapping is a serial method requiring point-by-point collection of diffraction patterns as a nanobeam scans the field of view (FoV), and acquisition requires tens of minutes for each FoV. For *in situ* studies of grain growth in nanocrystalline thin films, thousands of images must be collected in real time for a single experiment. It is thus necessary to use parallel-beam BF imaging, where acquisition times can be <1 s per frame on a standard charge-coupled device (CCD) camera. Manual segmentation of this many images, however, is far beyond the practical limitations of hand tracing, and so progress on these experiments necessitates automated grain boundary detection in BF images, which is addressed in this study.

While conventional edge detection (Spontón and Cardelino, 2015) fails, deep learning (DL) models in the form of convolutional neural networks (CNNs) have shown broad success in microstructural image analysis due to their ability to learn salient features at both coarse and fine scales that can ultimately be used for image or pixel classification (Holm et al., 2020). Unfortunately, supervised learning approaches for grain boundary identification in BF TEM are generally bottlenecked by limited available training data due to the expense of image micrographs often require fewer images than training for more general computer vision tasks, likely due to their elevated information density (Holm et al., 2020).

The U-Net architecture captures salient features at fine and coarse scales and is thus a good candidate model for this application. It consists of a contracting sequence of learned convolution filters (encoder) for learning salient image features at multiple scales, followed by an expanding sequence of learned convolutions (decoder) that translates the learned features into localized pixel classification. This leads to a grayscale segmentation that retains the same resolution as the input image, which can then be thresholded or subject to more complex postprocessing.

In this work, a supervised DL approach using the U-Net architecture is coupled with a significant corpus of expertly verified BF image/hand-tracing pairs and a carefully constructed postprocessing scheme to reconstruct polycrystalline microstructures from BF TEM images of Al thin films. The model is validated via direct comparison of the microstructures as reconstructed by the U-Net model to the microstructures as reconstructed by hand tracing. Then, the technique is tested on newly collected images of an Al thin film recorded on a modern CCD camera, demonstrating its applicability to newly acquired data without the need for extensive hand tracing. Finally, the application of the model is shown to have promising results on data collected during *in situ* heating experiments on Pt and Pd thin films. Taken together, the results suggest that the training data introduced here provide a solid foundation for supervised learning approaches such as U-Net to learn the salient features of BF TEM micrographs of polycrystals and identify grain boundaries, with the potential for generalizability to a range of materials and microstructures, opening the door to high-throughput, rapid analysis of large and *in situ* data sets.

## Materials and Methods

### Training, Validation, and Test Data

Studies from the early 2000s generated a sizable quantity of hand-labeled micrographs, the microstructural metrics for which are summarized in the article by Barmak et al. (2013).

These were originally used to analyze geometric and topological features of polycrystalline thin-film microstructures. Collecting these data was a time- and labor-intensive process, but they are now an invaluable asset, forming the basis of the training and validation data for the proposed supervised learning approach. The data were collected and annotated as follows. A set of 100-nm-thick Al thin films were sputter deposited and imaged in BF mode at various magnifications on a Philips EM420 microscope. Details of film deposition, annealing, and imaging are given in Supplementary Section 1. For each FoV, one to five images were taken, varying the sample tilt angle between −1° and 1° to change the diffraction conditions and increase the number of identifiable boundaries while minimizing image distortion. Large tilt angles lead to foreshortening, as the projection of the sample onto the image is scaled along the transverse direction and can lead to more stage movement and consequently more difficulty in aligning the images; as such, tilt angles are conventionally chosen with magnitudes ≤ 1°. Grain boundaries were manually identified by tracing enlarged photographic prints using a marker on transparencies; all tracings were verified by a second human annotator, such that no boundary or grain was included in the final set unless both annotators agreed that there was indeed a boundary or grain in the traced location. In total, 40 image-tracing pairs survive for these analyses, corresponding to 117 total BF images. The ground truth training data sets were constructed from 12 of these image-tracing pairs (39 BF images), and the model was validated using the remaining 28 FoVs (78 BF images). A representative image with its hand tracing overlaid can be seen in Figure 2a.

To construct the training data set, the images at each tilt from 12 FoVs (39 images) were aligned to their associated hand tracing using an in-house software package; any associated rescaling was accounted for in subsequent measurements. The original resolution of the scanned BF images was 4,096 × 4,096 px (see Supplementary Section 2), but after alignment and cropping, they were rescaled to match the resolution of their respective hand tracings, which were scanned in the early 2000s. This resulted in variable resolutions between 3,758 × 3,758 and 5,846 × 5,846 px. The hand tracings did not include boundaries that intersected an edge or regions where there was ambiguity in the image, as can be seen in the regions labeled (1–3) in Figure 2a. To exclude these areas from training, the pairs were cropped into 299 total patches of BF images and associated hand-traced boundaries. The size of the regions was variable, for two reasons: first, this maximized the number of reliable boundaries in a given region; second, variable patch sizes provide a more robust training data set, with boundaries and grains at various scales relative to the training image patch. Two such examples can be seen in Figures 2b and 2c. These pairs were then resized to a resolution of 256 × 256 px using an antialiasing method in the scikit-image© package (resize), to reflect the U-Net's input layer resolution. The remaining images were used as validation data. All hand-traced data were processed into binarized skeletons and analyzed in ImageJ (Schindelin et al., 2012; Schneider et al., 2012) to extract grain areas and centroid locations.

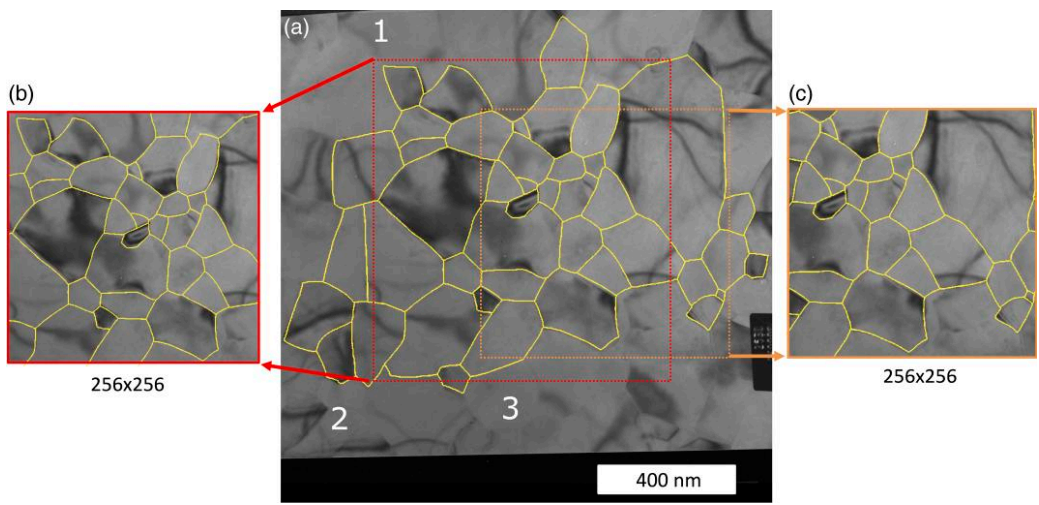

**Fig. 2.** (**a**) Scanned and inverted photographic negative of a representative BF TEM image of a 100-nm-thick film, with its hand tracing overlaid . (**b,c**) Patches used for training, which were taken from this FoV. Before U-Net training, the patches were rescaled to a resolution of 256 × 256. They are shown with their corresponding patch of hand tracing overlaid. Labels 1, 2, and 3 have been placed near regions with ambiguity to the presence of a grain boundary.

Next, to demonstrate that the supervised learning approach trained and validated on the data described above can be effectively used to extract microstructural information from data collected from new samples, a new 100-nm-thick Al film was sputter deposited, annealed, and subsequently imaged on an FEI Talos F200X S/TEM in BF TEM mode. This data set is referred to later as "test" data. Details of film preparation, annealing, and imaging are given in Supplementary Section 3. Each FoV was imaged at sample tilts of −1°, 0°, and +1° to maximize the number of identifiable boundaries; image shifts were partially corrected by stage translation, but alignment was still required before segmentation using the trained model. This was performed using an in-house software package, with more details in Supplementary Section 6. Three tilts from 49 FoVs were captured from the film for four annealing conditions, yielding in total 147 images. One such FoV's tilt series can be seen in Figures 3a–3c. Measurement of all microstructures was performed in ImageJ (see Supplementary Section 5).

Finally, to illustrate the transferability of the model to data collected on materials other than Al, even using a different imaging modality, a 50-nm Pt film was sputter deposited on a Protochips (Morrisville, NC, USA) MEMS heating chip. Imaging was performed on the FEI Talos F200X S/TEM in conical BF (CBF) mode in a Fusion Select™ heating holder using the Protochips AXON™ software for drift correction and data management. Here, to vary the contrast, the beam is tilted and rotated rather than the sample; this eliminates the abrupt image shift associated with sample tilting, reducing the contrast variation but rendering *in situ* data collection possible. The images were captured at approximately 1 frame per second. Before the CNN-base grain boundary detection was performed, a nonlocal Bayesian denoising algorithm (Lebrun et al., 2013) was applied to account for noise introduced by short integration time necessitated by live acquisition. More experimental details about the deposition and imaging conditions can be found in Supplementary Section 3.

## DL-Based Grain Boundary Segmentation Using U-Net

The automated grain boundary detection process was performed using a supervised DL approach based on the U-Net architecture. As noted earlier, U-Net is a CNN that was first developed for biomedical image segmentation (Ronnenberger et al., 2015). These images face similar problems to the grain boundary detection problem, including complex image characteristics, like intrafeature contrast and touching objects of the same class, and training data that are sparse and laborious to generate. Given the similarities in the images of interest, the U-Net architecture is an attractive option.

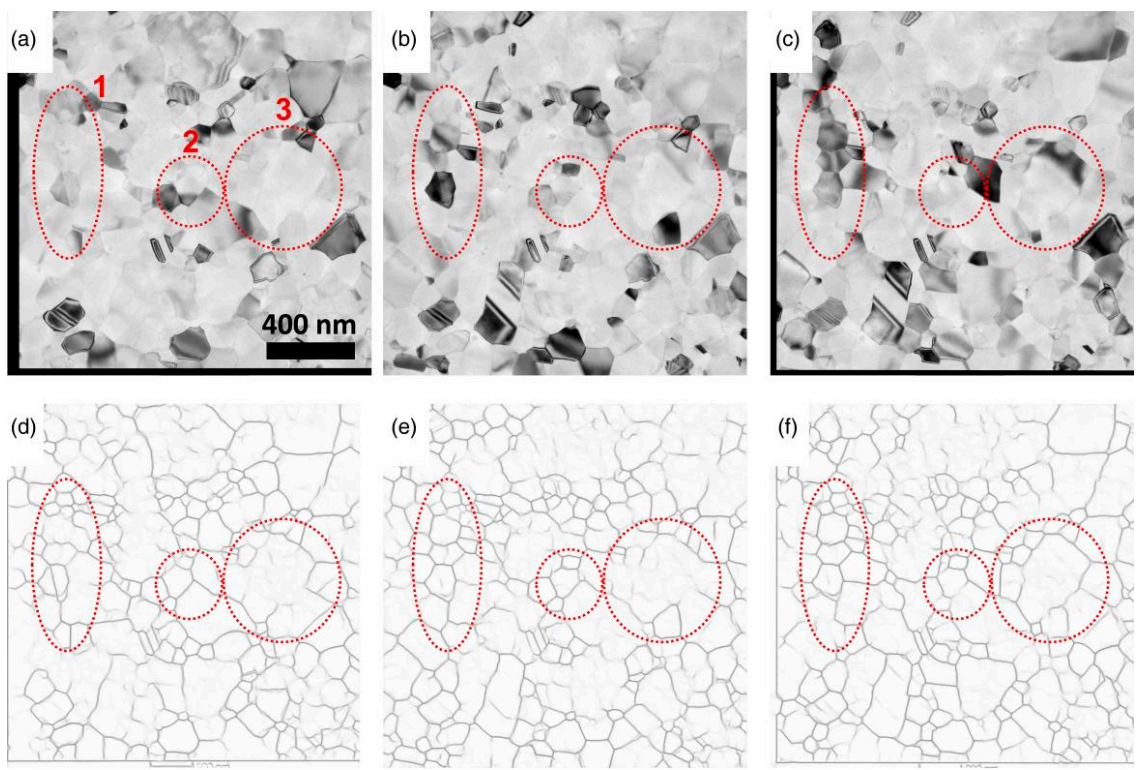

**Fig. 3.** BF TEM images and the trained U-Net model's grain boundary predictions of a recently deposited 100-nm-thick Al film, captured at a sample tilt of (**a,d**) −1°, (**b,e**) 0°, and (**c,f**) + 1°. The ±1° tilted images were aligned to the 0° image; shifts due to image alignment are visible in the black regions near the edges of **a** and **c** and the artificial boundaries identified around the edges of **d** and **f**. Features labeled 1 and 2 in red are regions where diffraction contrast (**a–c**) variation leads to significant differences in the visibility of grain boundaries (**e–f**). Feature 3 corresponds to a likely bend contour in a large grain. BF images were captured at 200 kV with a 30-μm objective aperture.

As in Ronnenberger et al. (2015), the contracting feature extraction sequence consists of five layers of $3 \times 3$ convolutions followed by the ReLu activation function (Goodfellow et al., 2016), where downsampling is performed between layers by a $2 \times 2$ max pooling operation with a stride of 2 and the number of filters is doubled at each subsequent layer. This is followed by an expanding localization sequence that ultimately leads to pixel classification, only differing from the contracting sequence in that the max pooling is replaced by upsampling. A number of convolution filters are halved at each layer, and each layer is concatenated with the corresponding feature extraction output.

For this work, the model was trained for 66 epochs. At each epoch, the 299 pairs of training patches were randomly augmented. To maintain the physical traits of the BF TEM data, the augmentation consisted only of rigid 90° rotations and horizontal and vertical reflection; this excludes image stretch and shear that are present in biological samples and included in the original implementation (Ronnenberger et al., 2015). Before training, 10% of the pairs were randomly selected and used only as a training validation set to measure the progress of the training at each epoch. The output at each epoch is subject to a sigmoid activation function (Goodfellow et al., 2016), and the model weights are updated to minimize an unweighted cross-entropy loss function that compares the model predictions and the ground truth hand-labeled grain boundaries in the randomly selected image-tracing pairs (Ronnenberger et al., 2015).

For each input image, the U-Net model outputs an 8-bit grayscale image of predicted boundary locations. In typical segmentation problems, a single threshold is usually sufficient to produce a binarized image suitable for further analysis. In contrast to typical foreground–background semantic segmentation problems, however, grain boundary segmentation involves much finer features of interest. Rather than distinguishing one or multiple objects from a solid background, the goal is to identify narrow "foreground" lines amid a "background" with complicated, inhomogeneous contrast variation. To accurately preserve the fine boundaries while minimizing oversegmentation, special attention is required during postprocessing, especially in binarization.

Following the general postprocessing approach of Carpenter et al. (1998), for each FoV, the grayscale U-net predictions are combined from the several tilts using a logical OR operator; this composite grayscale image then is post processed using double thresholding, dilation, hole closing, skeletonization, and pruning. The grayscale upper and lower thresholds, dilation size, minimum hole area, and pruning distance are determined empirically, as described in the Determining Postprocessing Parameters section. Figure 4 summarizes the entire image processing pipeline, with BF images of a new AI film (a–c) and the U-Net output after each stage of post processing (d–l).

### Determining Objective Functions for Model Validation

Even to expert human annotators, there is some ambiguity in the exact boundary positions in BF images of nanocrystalline films. This makes quantitative comparison of the results for the model to the hand tracings especially challenging. Simple metrics, like image differences, are too sensitive to small deviations in boundary positions. The goal of automated grain boundary tracing, however, is to robustly quantify important microstructural parameters, such as grain size, grain shape, number of neighbors, and other geometric and topological metrics. To that end, if the U-Net tracing reconstructs a microstructure that is similar enough to the microstructure obtained by expert hand tracing, then it can be considered successful. To the first order, then, two key aims of the model are to identify grains in the same locations as a human observer and to find the same grains in a given FoV. From this perspective, the hand-traced images are regarded as a ground truth for subsequent image comparison.

To simplify the problem and avoid sensitive pixel-wise comparisons, the centroid locations of identified grains are considered instead, effectively reducing each reconstructed microstructure to a point process (Diggle, 2021), while preserving relevant physical length scales. Using ImageJ (Schindelin et al., 2012), grain centroids are identified in the microstructures generated by hand tracing (reference data) and the microstructure generated by the U-Net model (comparison data). After uniformly shifting the comparison centroids so that their collective center of mass was aligned to that of the reference data, centroid pairs corresponding to the same grain in the two microstructures are identified by finding the closest points in the two images such that the distance between a pair of points is less than a two-dimensional cutoff distance $R$ (taken here as $0.3r_{nn}$, where $r_{nn}$ is the mean intercentroid distance). The resulting mapping is one to one with an associated distribution of intercentroid distances. After this pairing process, some centroids may remain unpaired in either set and will be referred to hereafter as "orphans."

To assess the quality of the match between the two sets of centroids, two objective functions, $\Omega_i$ ($i = 1, 2$), with $0 \le \Omega_i \le 1$ with $0$ (1) corresponding to the best(worse)-case scenario are considered here. Taking $\{r_i^h | i = 1, 2, \ldots, N^h\}$ as the set of $N^h$ hand-traced centroids and $\{r_i^u | i = 1, 2, \ldots, N^u\}$ as the set of $N^u$ U-Net determined centroids, the first objective function is given by Equation (1):

$$\Omega_1 = \frac{1}{NR} \sum_{i=1}^{N} \left| \overline{r}_i^u - r_i^h \right|, \tag{1}$$

where the overbar denotes the minimum distance, U-net centroid that matches the corresponding hand-traced centroid using the procedure outlined above, the double bars denote the 2-norm (i.e., Euclidean distance) of a vector, and $N$ is the number of pairs of matching centroids. Thus, $\Omega_1$ is a disregistry function that measures the degree of matching between the collection of paired points. This metric can be thought of as analogous to the (normalized) Wasserstein distance (Panaretos & Zemel, 2019) for two thinned point processes (Diggle, 2021; Baddeley et al., 2016).

The second objective function, $\Omega_2$, measures the degree to which the two microstructures identify the same grains and is computed as the fraction of all identified centroids that are orphans by Equation (2):

$$\Omega_2 = 1 - \frac{2N}{(N^h + N^u)}, \quad N \le \min(N^h, N^u). \tag{2}$$

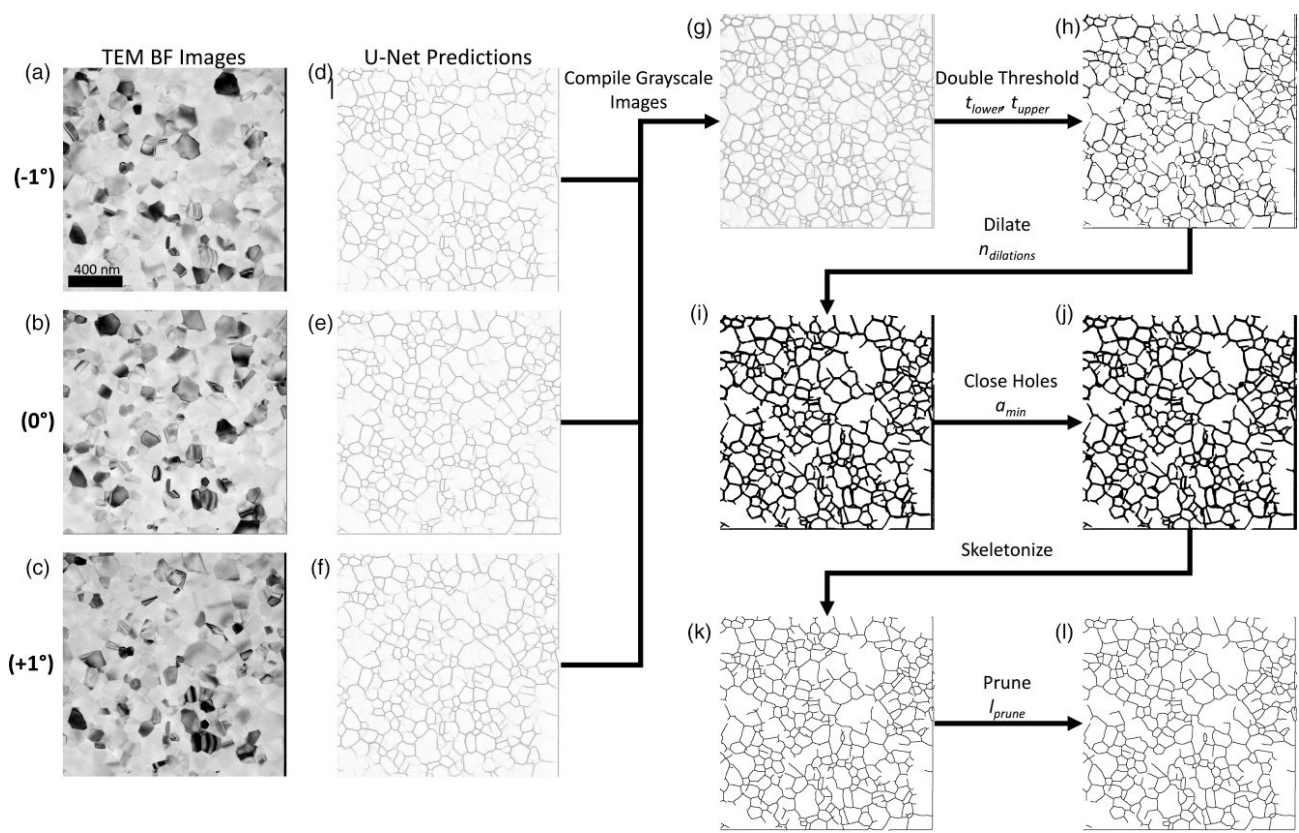

**Fig. 4.** Complete U-Net flow used for all images, with images of a recently deposited 100 nm-thick Al film as an example. Three aligned BF images captured at sample tilts of (**a**) −1°, (**b**) 0°, and (**c**) +1°. (**d–f**) The corresponding grayscale U-Net predictions. Postprocessing results after (**g**) logical OR combination, (**h**) double thresholding with an upper threshold of 213 and a lower threshold of 183, (**i**) three dilations, (**j**) hole closing for holes smaller than 60 px in area, (**k**) skeletonization (boundaries are dilated in the figure for visibility), and (**l**) pruning with prune length 70 px (boundaries are dilated in the figure for visibility). Black borders are an artifact due to the image shifts associated with sample tilting that appear as straight-line artifacts along the edges; these are cropped before analysis.

The orphan fraction $\Omega_2 = 1$ would indicate that all centroids are orphans and there are no matching pairs whereas $\Omega_2 = 0$ would indicate that no centroids are orphans (i.e., every centroid is paired). These two objective functions measure how well the U-Net tracings match the hand tracings and the degree to which a given image is over- or undersegmented by U-Net. They comprise the basis of a multiobjective optimization scheme to select postprocessing parameters, as described below. It is noted that a related approach combining similar expressions into a single objective function was developed by Schuhmacher and Xia (2008).

### Determining Postprocessing Parameters

The U-Net model outputs an 8-bit grayscale image, but measurements must be performed on binary images. Obtaining a usable image from the U-Net predictions requires several postprocessing steps, each with an associated parameter or set of parameters. While an end-to-end method would be preferred, retaining a separate postprocessing step yields a more flexible approach for data sets unlike the original training data. The three relevant steps are as follows: (1) double thresholding, with a lower threshold value of $t_L$ and an upper threshold value of $t_U$; (2) dilation, with $n$ dilations; hole filling, with a minimum grain area of $a_{min}$; and (3) pruning, with a prune distance of $l_{prune}$ (see Fig. 4). For a given data set, these parameters must be determined in a systematic and reproducible way.

To accomplish this for the training and validation data, the model was applied to the full BF images of the FoVs (not the patches) used during training, including all tilts for each of the 12 FoVs, totaling 39 images. For each training FoV, a large number (~6,000) of postprocessed images were generated, varying the postprocessing parameters in each image to sample the entire parameter space. While a more modern machine learning approach might be preferable, as of now, loss functions like intersection over union and others are not well adapted for fine features like grain boundaries, or FoVs where objects of interest fill virtually the entire space. To select the optimal parameters, the objective functions defined in the Determining Objective Functions for Model Validation section were calculated for each postprocessed image by comparing the processed U-Net output to the corresponding hand-traced microstructure. For a given FoV, the parameters that generated an objective function pair ($\Omega_1$, $\Omega_2$) with the minimum vector magnitude were recorded. The thresholding and dilation parameters were then averaged across all FoVs. The objective functions proved generally insensitive to the minimum area $a_{min}$ and the prune distance $l$. These parameters were therefore anchored to characteristic microstructural dimensions, with $a_{min}$ taken to be 1% of the estimated mean grain area and $l$ is taken to be the perimeter of this minimum allowed grain as reasonable physical limits for the minimum size of a real grain and length of a real boundary.

In the postvalidation problem considered here, the values of objective functions (i.e., orphan fraction and disregistry) reflect the degree of agreement between a U-Net tracing and a hand tracing. These values may be presented in the form of a two-dimensional objective function space as visualized in Figure 5. It shows the objective function space for every combination of postprocessing parameters for all training FoVs; each color corresponds to a different FoV. This plot evinces a Pareto front (Jin & Sendhoff, 2008), highlighting the trade-off inherent in simultaneously optimization of the disregistry and orphan fraction. In this context, the agreement is quantified by multiobjective optimization (i.e., minimization of the distance to the origin). As the objectives are often in conflict with one another, improving one often results in the degradation of another and so there is typically no single optimal solution. A Pareto front is a collection of solutions in the objective function space that embodies these trade-offs (Chong and Zak, 2013; Akbari et al., 2014). Thus, the optimal operating conditions for a given FoV fall along this front, which is approximately drawn to guide the eye in Figure 5. Because the $t_U$, $t_L$, and $n_{dilation}$ used are averages of the best-performing images for the tested FoVs, the true operating conditions used during validation fall inside the front in Figure 5. This choice optimizes the performance on the ensemble of images rather than for any one FoV.

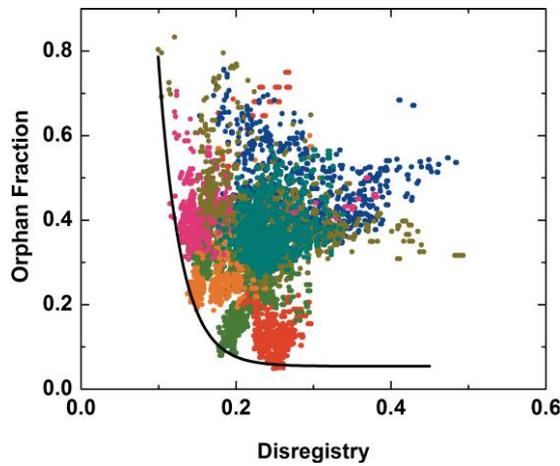

**Fig. 5.** Objective functions for all training FoVs at all post processing parameter, totaling 48,308 test images. Each color represents a different FoV. An approximate Pareto front is drawn as a solid black line to guide the eye.

To determine the postprocessing parameters for the recently acquired test data, some amount of ground truth data were required for comparison. To reduce the work needed to label boundaries, the initial U-Net prediction was used as a starting point for four FoVs. A human observer completed missing boundaries and eliminated boundaries that were spurious to generate a "machine-assisted hand tracing" using the application Notability on an Apple iPad Pro and with an Apple Pencil. These machine-assisted tracings were exported as *.png images and scaled to 4,096 × 4,096 px, the same resolution as the captured BF micrographs. The resolution of these tracings was not a critical factor, as the measurements of the objective functions was done based on physical units (nm), and so only the dimensions of the FoV need be preserved to accurately measure the microstructures. The mean grain size was estimated from the machine-assisted tracings and used to determine the minimum grain area and the prune length. The procedure above was followed to determining the thresholds and number of dilations for these particular image characteristics.

## Results and Discussion

### Validation Set: AI Images from Prior Studies

To validate the results of the trained model and post processing approach, it was applied to the remaining 28 FoVs (78 BF images) available from prior studies that were not used during training. These images were collected under the same conditions as the training data and have hand tracings (ground truth) for comparison. The U-Net predictions were postprocessed using the parameters determined in the Determining Postprocessing Parameters section from the training data set. First, the U-Net microstructures are compared to the hand traced ground truth using the multiobjective optimization scheme outlined above. Then, population-level statistics, such as the mean grain size and grain size distributions obtained from the two measurement techniques are compared. In all cases, it is observed that both techniques reproduce microstructures that are sufficiently similar.

To quantify the statistical significance of the results, the grain centroids of the hand-traced grains are approximated as $N$ centers of fixed, mutually independent circles with each circle having a radius of $R = 0.3 r_{nn}$, where $r_{nn}$ is the mean intercentroid distance for the hand-traced reference microstructure. From this perspective, each of the paired U-Net centroids is contained in exactly one such circle, and these centroids can, taken together, be regarded as a collection of independent, identically distributed random variables. $\Omega_1$ is therefore a point estimator of the normalized mean separation, $\mu/R$, between the U-net and the hand-traced centroids.

To validate the results, this mean separation $\mu/R$ is compared to the distance that would be obtained if the U-Net centroids are randomly distributed in the aforementioned circles; if it is much closer to zero, then it follows that there is a reasonably good match between the microstructures. To accomplish this, a hypothesis test is performed; the null hypothesis is taken as $H_0 : \mu/R = \mu_0/R$, against the alternative hypothesis $H_A : \mu/R < \mu_0/R$, where $\mu_0$ is the mean distance for a spatially uniform distribution of points in a circle of radius $R$. To obtain $\mu_0$, it is noted that the probability density function for the uniform distribution of points is given by Equation (3):

$$p(r) = \frac{2r}{R^2} \, \Theta(R - r),$$  (3)

where $\Theta$ denotes the step function. Since the expected value $E(r^n) = 2R^n/(n+2)$, it follows that $\mu_0/R = 2/3$ and $\sigma/R = 1/\sqrt{18}$, where $\sigma$ is the standard deviation of the intercentroid distance.

A one-tail hypothesis test (Ross, 2009) for the case that $N \geq 30$ given $\sigma$ begins with the calculation of the z-score, given by Equation (4):

$$z = \frac{\Omega_1 - \frac{\mu_0}{R}}{\left(\frac{\sigma}{\sqrt{N}R}\right)},$$  (4)

and determination of the associated p-value. Taking $z$ to be normally distributed, one finds that for $\Omega_1 = 0.17$, as obtained from the best-performing position on the Pareto front for the training data, one finds that the value of the statistic $z = -11.6$ for $N = 30$ and that the associated $p \approx 0$, and so one can reject $H_0$ in favor of $H_A$. This $\Omega_1$ value and number of pairs is typical across the training and validation data, with the mean disregistry across the validation set being $0.21 \pm 0.01$ and a corresponding vanishingly small $p$-value. The worst performing FoV had a disregistry value of $\Omega_1 = 0.44$. With a small number of pairs $n = 6$, $z = -2.4$ giving a $p$-value of 0.02, still clearly allowing the rejection of $H_0$ in favor of $H_A$. Thus, there is sufficient evidence that the U-net centroids are (much) closer to the hand-traced centroids than would be expected by chance, suggesting that, by this measure, there is a good match between the U-net and the hand-traced microstructures. It is noted that in the cases where one can calculate a $t$-statistic in place of Equation (4) with an estimate of $\sigma$ from the data and the associated $p$-value from the Student's $t$ distribution (Ross, 2009), the results remain essentially unchanged.

While the centroids are very unlikely to be spuriously identified as pairs, some grains are identified in one technique but not the other and remain orphaned. One source of these differences lies in areas with challenging-to-identify boundaries; here, the human annotators often opted to exclude these regions so that later measurements were only made of grains with boundaries identified with high certainty. The U-Net model can make no such judgment calls, and so it predicts boundaries in areas that human observers chose not to trace. In other FoVs, U-Net finds many fewer grains than hand tracing, reflecting a set of postprocessing parameters optimized on the ensemble rather than for each FoV. To demonstrate that the extra grains identified by U-Net are not entirely spurious and do not significantly degrade the quality of measured microstructural properties, it is of interest to determine whether the U-Net microstructure is similar to the hand-traced microstructure at the grain population level. To assess this, the grain sizes and grain size distributions are measured from the full set of grains identified by both techniques and these statistics are then compared. In all cases, ImageJ (Schindelin et al., 2012) was used to analyze the skeletonized grain boundary networks. Figures 6a–6c show a representative BF TEM image with its hand tracing overlaid and the BF TEM image with the U-Net tracing overlaid. Figures 6d and 6e show the reduced grain size distributions obtained from U-Net and from hand tracings. These distributions are fit to log-normal probability density functions, which are shown as solid curves. Log-normal $Q$–$Q$ plots are inset and show a generally good fit for both distributions, with small deviations in the U-Net microstructures only in the left tail of the distribution.

Following Carpenter et al. (1998), the first and second moments of the distributions are also considered. The equivalent circle diameter of mean grain area, $d_{<A>}$, for the U-Net and the hand-traced microstructures are $165 \pm 8$ nm and $170 \pm 8$ nm for 1,359 and 1,429 identified grains, respectively. Similarly, the mean diameters $<d>$ were found to be $148 \pm 7$ nm and $153 \pm 7$ nm for the same set of grains. Here, the error bars were determined using the results from Carpenter et al. (1998). When the normally distributed logs of the diameters are considered, the Student $t$-test shows that the mean values agree at the 95% confidence level, with a $p$-value of 0.70. These diameters are summarized in Table 1.

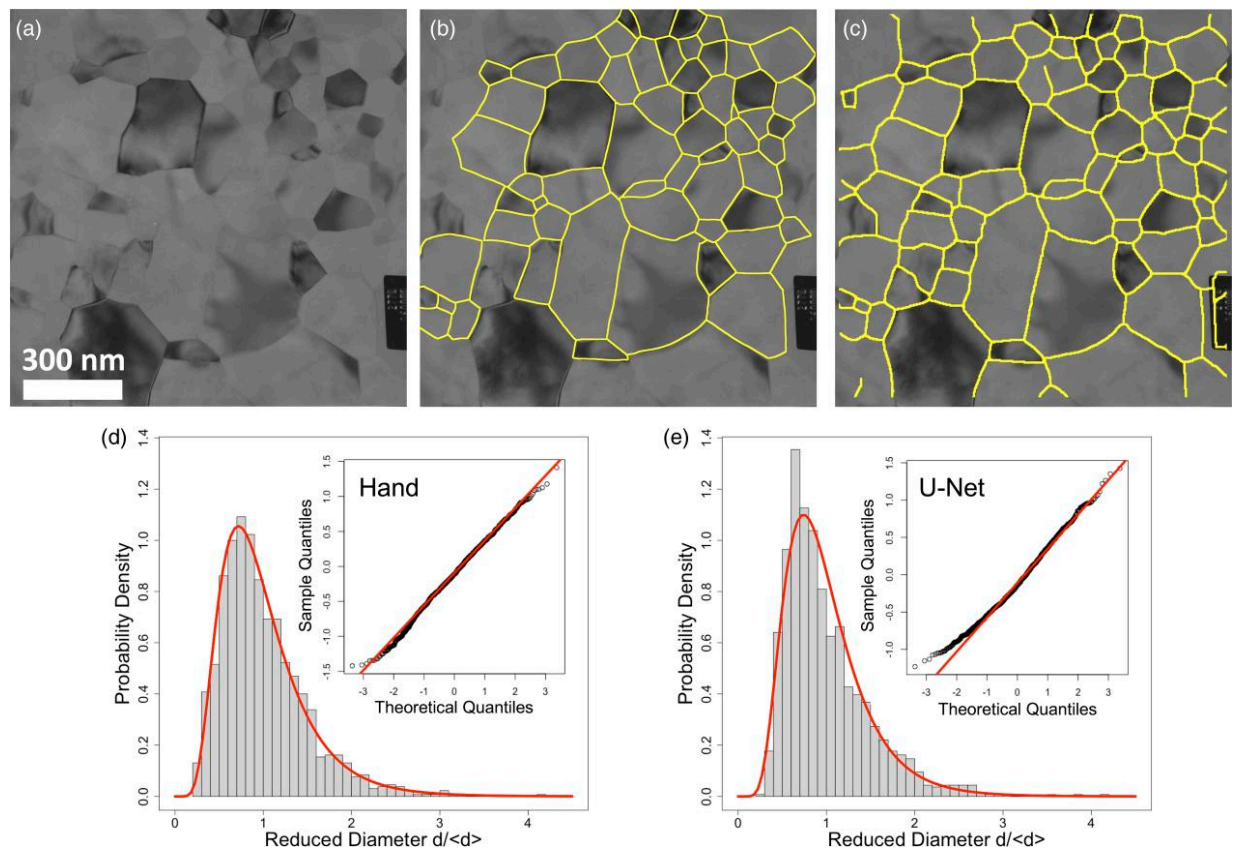

**Fig. 6.** (**a**) Representative BF TEM image of an image used for validation with its (**b**) hand tracing and (**c**) postprocessed U-Net output overlaid in yellow. Boundaries are dilated for clarity. Grain size distributions obtained from the validation data via (**d**) hand tracings and (**e**) U-Net measurements. Log-normal fits are shown as solid curves. Insets: Corresponding normal $Q$–$Q$ plot for the logs of the reduced diameters with a normal distribution shown as a solid line.

**Table 1.** Mean Grain Diameters from Hand Tracing and U-Net.

| Data Set | Technique | $d_{<A>}$ (nm) | $<d>$ (nm) | $N$ (Grains) |
|---|---|---|---|---|
| Validation | Hand tracing | $170 \pm 8$ | $153 \pm 7$ | 1,429 |
| | U-Net + postprocessing | $165 \pm 8$ | $149 \pm 7$ | 1,359 |
| Newly deposited (test) | Machine-assisted hand tracing | $113 \pm 6$ | $101 \pm 5$ | 813 |
| | U-Net + postprocessing | $114 \pm 6$ | $99 \pm 5$ | 7,398 |

**Table 2.** *P* Values for Statistical Tests Performed, Comparing Results Obtained from U-Net and Postprocessing to Ground Truth Data.

| | Statistical Tests for Distribution | | | Statistical Test for Centroids |
|---|---|---|---|---|
| Sample | Cramér–von Mises | Kolmogorov–Smirnov | $t$ log($d$) | Disregistry |
| Validation | 0.16 | 0.20 | 0.70 | <0.01 |
| Newly deposited (test) | 0.22 | 0.16 | 0.84 | <0.01 |

To compare the grain size distributions obtained from the two measurement techniques with less sensitivity to noise in the tails than, for example, the *F*-test, cumulative distribution functions are considered for the distributions of reduced effective grain diameters, $d/<d>$, using a Kolmogorov–Smirnov (KS) test (Press et al., 1992) and a Cramér–von Mises (CVM) test (Andersen et al., 1993). The result of the KS test yields $p = 0.204$, and the CVM test yields $p = 0.157$, indicating in both cases that there is insufficient evidence to suggest that the distributions are different. These *p*-values are summarized in Table 2.

Taken all together, the results presented above indicate that U-Net predictions for microstructures based on BF TEM images are sufficiently close to gold-standard hand-traced microstructures in multiple respects: at the level of the locations of individually identified grains, it is observed that the positions of grain centroids are significantly closer than expected for a random case; at the grain population level, for metrics like grain size and grain size distribution, there is likewise significant overlap which cannot be attributed to chance. This result invites the application of this U-Net to investigate samples for which a body of hand tracings is not available.

### Test Set: Digital Images of New Al Films
When the U-Net model was applied to newly acquired images of a 100-nm-thick Al thin film, despite notably different image characteristics to the training and validation data, the grayscale U-Net predictions such as those presented in Figures 3d–3f and 4d–4f are very promising for microstructural interrogation. It should be noted that U-Net predictions for these images were performed at a resolution of $1,024 \times 1,024$ based on an approximation that there were twice as many grains per FoV in the test data as the training data. The regions labeled 1 and 2 of Figure 3 illustrate the variable contrast from tilting, as well as the changing U-Net predictions in this region. Furthermore, the prediction does not include any false boundaries in the region labeled 3 where there is a visible bend contour.

Using the grayscale U-Net images and postprocessing parameters as determined from four machine-assisted hand tracings (Determining Postprocessing Parameters section), excellent grain boundary segmentation is achieved. Figure 7a shows a representative FoV with its measured grain boundary network overlaid in red. The grain size distribution was measured for this Al film using the U-Net outputs from 49 FoVs and can be seen in Figure 7b. In total, 7,398 grains were automatically detected in the final microstructures, representing an enormous leap in throughput from what is possible via hand tracing alone. The distribution of their diameters is well described by a log-normal density function, and a fitted log-normal probability density function is shown as a solid red curve in Figure 7b. A log-normal normal *Q–Q* plot of the logs of the reduced diameters is inset in Figure 7b and illustrates the log-normality of the distribution. The empirical distribution only departs from log-normality in the left tail of the distribution, as was observed in the validation data set.

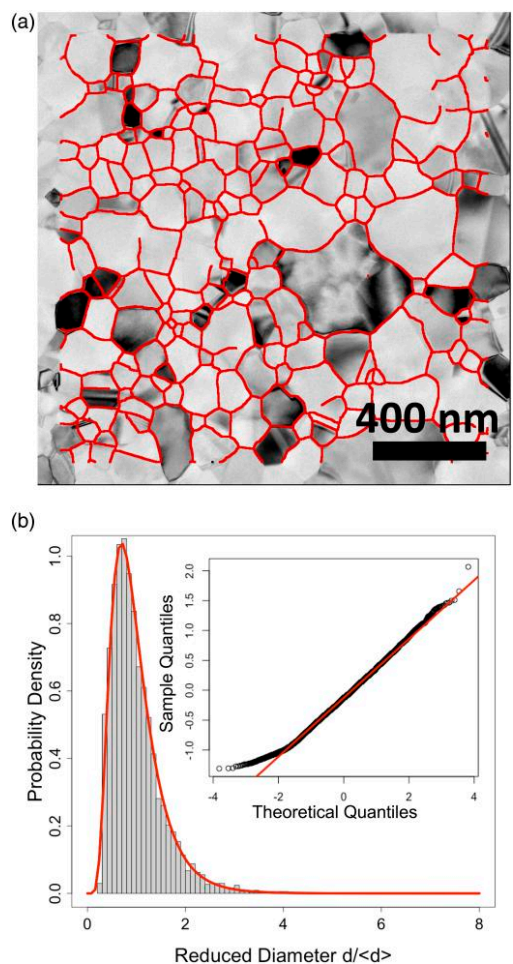

**Fig. 7.** (**a**) A BF TEM image of the recently deposited 100-nm-thick Al film that had been annealed at 185°C, with its postprocessed U-Net grain boundary network overlaid in red. Boundaries are dilated for clarity. The edges are cropped to account for artifacts that result from the alignment of the tilt series. (**b**) Probability density of reduced diameters for 7,398 grains measured from this film's postprocessed U-Net grain boundary networks. The log-normal fit of the histogram is shown as a solid red curve. Inset: A normal *Q–Q* plot of the logs of the reduced diameters with a normal distribution shown as a solid red line.

Comparing the four microstructures obtained from the automated method to the corresponding four machine-assisted hand tracings, using the hypothesis test described in the Validation Set: Al Images from Prior Studies section, the worst performing FoV had a value of $\Omega_1 = 0.49$ for 145 pairs, yielding $p \approx 0$, allowing the rejection of the null hypothesis and conclusion that the matched points are significantly closer than random. At the grain population level, the film microstructure measured using the automated technique had an equivalent circle diameter of mean grain area, $d_{<A>}$, of 113 $\pm$ 6 nm ($N = 7{,}398$), compared to the 114 $\pm$ 6 nm ($N = 814$) obtained in from the machine-assisted hand tracings. The diameters are summarized in Table 1. Comparing the distributions, a *t*-test shows that the mean circle equivalent grain diameters agree, with a *p*-value of 0.84; similarly, the KS test and CVM test show that there is not sufficient evidence to suggest the cumulative distribution functions of diameters disagree, with respective *p*-values of 0.16 and 0.22 (see Table 2 for comparison to other values obtained.)

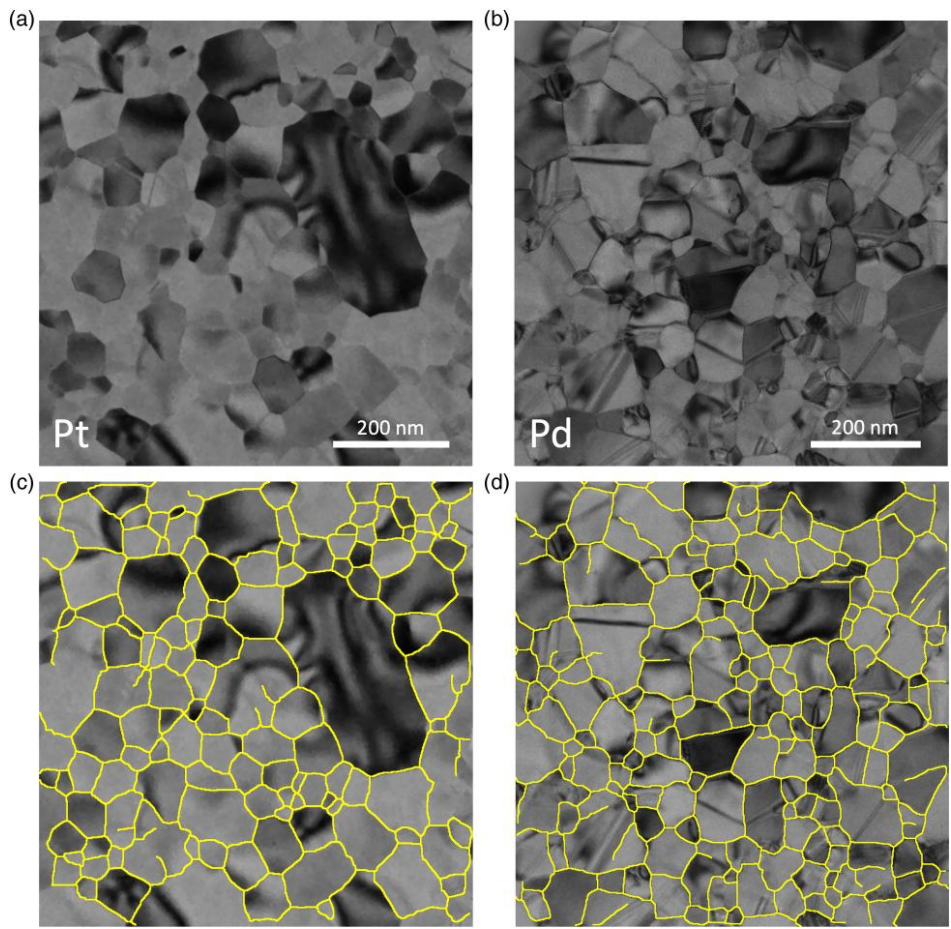

**Fig. 8.** (**a**) A BF TEM image of a 50-nm-thick Pt film at 800°C during an *in situ* heating experiment. (**b**) A BF TEM image of a 30-nm-thick Pd film at 400°C during an *in situ* heating experiment. (**c,d**) Images from **a** and **b** with grain boundary segmentations overlaid, generated from three diffraction conditions for Pt and one diffraction condition for Pd. They were generated using the same model weights as the Al data, and the postprocessing was optimized in the same way as the Al data, based on one machine-assisted hand tracing for each material.

The grain size distributions observed here are also qualitatively consistent with the universal grain size distribution observed in metallic thin films over years of data collected on Al and Cu films (Barmak et al., 2013), obtained from hand tracing, gradient-based automated tracing (Carpenter et al., 1998), and PED orientation mapping (Liu et al., 2014; Rohrer et al., 2017). This reiterates the reproducibility of the experimental distribution obtained independent of film thickness, annealing time, and annealing temperature and emphasizes the independence of the results with respect to the method of grain size determination. It should finally be noted that when the automated technique is applied with suboptimal resolutions and/or postprocessing parameters, the results do not reliably replicate these findings. Generally, both of these can be optimized using the objective functions and Pareto-type approach presented in the Determining Postprocessing Parameters section, although these values are less sensitive to changes in resolution than to changes in postprocessing.

The U-Net model and associated postprocessing scheme's success on a data set unlike that on which the CNN was trained, with minimal additional manual input, point to an unprecedented flexibility for use in a host of microstructural analyses. The transferability renders it a powerful tool for grain boundary identification tasks in various microstructures under various imaging conditions. In particular, the results presented here represent a leap in throughput which opens the door to the analysis of the large data sets key to understanding microstructural phenomena, like the dynamics of grain growth.

### Transferability and Outlook

Of interest in any supervised machine learning approach is the generalizability to data that is not represented in the training set. The U-Net model trained on hand-labeled Al data from the early 2000s can be adapted with relative ease to new materials and imaging modes of polycrystalline materials. As an example, Figures 8a and 8b show BF images captured during *in situ* heating experiments performed on the 50-nm platinum thin film and 30-nm palladium thin film, respectively. One machine-assisted hand tracing was generated. Based on this labeling, the resolution of the outputted images for Pt (Pd) was chosen to be $672 \times 672$ px ($640 \times 640$ px), optimized via the objective functions defined in the Determining Objective Functions for Model Validation section.

As these data sets are small, containing 113 and 133 automatically detected grains for Pt and Pd respectively, they cannot be used for grain size distribution analysis; however, when compared to machine-assisted hand tracings, the disregistry (orphan fraction) values for Pt are 0.20 (0.18) and for Pd 0.24 (0.33), comparable to the results obtained from the Al data in earlier sections. While not unusually high, the larger orphan fraction in the Pd film results can partially be explained by the difficulty in identifying exceptionally narrow twins that have similar characteristics to bend contours; this is an issue that is the subject of ongoing work and can likely be resolved through additional training on data containing this type of microstructural features. This indicates that the technique introduced in this work shows great promise in analyzing these types of data and analyzing *in situ* data sets going forward. It also indicates that the training data set constructed from hand tracings is robust and unique in its depth, allowing the model to learn almost all salient features present in BF TEM images of nanocrystalline thin films.

## Summary and Conclusions

Collection of large-scale nanocrystalline thin-film microstructural data has historically required a proportionally largescale and labor-intensive manual identification of grain boundaries in BF transmission electron micrographs. This burden has partially been alleviated by the use of precession-enhanced electron diffraction-based orientation mapping techniques, but this method comes at the cost of a slow acquisition time, making the collection of data for very large numbers of grains expensive and rendering the technique unsuitable for *in situ* experiments. While conventional gradient-based techniques and CNNs have been applied in micrographs with simpler image characteristics, limited progress has been made in automating grain boundary identification in BF images, largely due to the complicated contrast arising from bend contours inherent to the samples and the physics of BF image formation.

The work presented here combines a carefully constructed training data set with a U-Net architecture and an optimized postprocessing algorithm (beyond the typical thresholding) to enable a supervised learning approach for grain boundary detection; together, the methods account for the challenging characteristics of the target data. The neural network was trained on a set of cropped Al BF micrographs and associated hand tracings; when the trained network was applied to the remaining images, the prediction successfully reconstructs microstructures consistent with those obtained from hand-traced images. Furthermore, when applied to newly generated data, given a careful selection of resolution and postprocessing parameters, the model successfully extracts new results. The automated approach shows good agreement with machine-assisted hand-traced microstructures and replicates decades of published work, where a universal grain size distribution has been repeatedly reproduced independent of measurement technique and other experimental parameters. Finally, the application of the model trained on hand-labeled Al data shows great promise for segmenting images captured during *in situ* heating experiments using CBF imaging of both platinum and palladium thin-film samples, even with no additional training of the neural network.

Taken together, these results indicate that the U-Net architecture is appropriate for microstructure studies in nanocrystalline thin films. Further, the ground truth training data set used in this study is robust and provides a strong platform for training neural networks to identify grain boundaries in BF TEM images, owing to its dense and high-quality manual segmentation that affords the trained neural network the ability to learn salient features of these types of images. In particular, the model is able to recognize faint, nearly invisible boundaries, while simultaneously ignoring bend contours and other image features, a pair of tasks which traditional algorithms fail to accomplish. This represents a leap forward for the analysis of nanocrystalline microstructure through BF TEM images, opening the door for future high-throughput, rapid analysis of large data sets at unprecedented time resolution.

## Availability of Data and Materials

All Al images and tracings used in this work are available from the Columbia Data Repository at https://doi.org/10.57783/w6n3-5b02. The modified U-Net and postprocessing implementations used to generate the results in this paper are available and can be downloaded from https://github.com/MatthewJPatrick/grain_unet. This code was based on the U-Net implementation originated at https://github.com/zhixuhao/unet.

## Supplementary Material

To view supplementary material for this article, please visit https://doi.org/10.1093/micmic/ozad112.

## Acknowledgments

The imaging work on the new Al samples was carried out in the Electron Microscopy Laboratory of the Columbia Nano Initiative (CNI) Shared Lab Facilities at Columbia University.

## Financial Support

This work was supported by the U.S. National Science Foundation (NSF) grant DMS-1905492 and the DMREF program under grants DMS-2118206 and DMS-2118197.

## Conflict of Interest

The authors declare that they have no competing interest.